\documentclass[conference]{IEEEtran}
\IEEEoverridecommandlockouts
\usepackage{cite}
\usepackage{amsmath,amssymb,amsfonts}
\usepackage{algorithmic}
\usepackage{graphicx}
\usepackage{textcomp}
\usepackage{multirow}
\usepackage{xcolor}
\def\BibTeX{{\rm B\kern-.05em{\sc i\kern-.025em b}\kern-.08em
    T\kern-.1667em\lower.7ex\hbox{E}\kern-.125emX}}
\begin{document}

\title{Practical experiences and value of applying software analytics to manage quality}

\author{\IEEEauthorblockN{Anna Maria Vollmer}
\IEEEauthorblockA{\textit{Fraunhofer IESE}\\
Kaiserslautern, Germany \\
anna-maria.vollmer@iese.fraunhofer.de}\\
\IEEEauthorblockN{Jari Partanen} 
\IEEEauthorblockA{\textit{Bittium}\\ 
Oulu, Finland \\
Jari.Partanen@bittium.com}
\and
\IEEEauthorblockN{Silverio Mart\'inez-Fern\'andez}
\IEEEauthorblockA{\textit{Fraunhofer IESE}\\
Kaiserslautern, Germany \\
silverio.martinez@iese.fraunhofer.de}\\
\IEEEauthorblockN{Lidia L\'opez}
\IEEEauthorblockA{\textit{UPC-BarcelonaTech}\\
Barcelona, Spain \\
llopez@essi.upc.edu}
\and
\IEEEauthorblockN{Alessandra Bagnato}
\IEEEauthorblockA{\textit{Softeam}\\
Paris, France \\
alessandra.bagnato@softeam.fr}\\
\IEEEauthorblockN{Pilar Rodr\'iguez}
\IEEEauthorblockA{\textit{University of Oulu}\\
Oulu, Finland \\
pilar.rodriguez@oulu.fi}
}

\IEEEoverridecommandlockouts
\IEEEpubid{\makebox[\columnwidth]{978-1-7281-2968-6/19/\$31.00~\copyright2019 European Union \hfill} \hspace{\columnsep}\makebox[\columnwidth]{ }}
\maketitle

\IEEEpubidadjcol

\begin{abstract}
\textit{Background:} Despite the growth in the use of software analytics platforms in industry, little empirical evidence is available about the challenges that practitioners face and the value that these platforms provide.
\textit{Aim:} The goal of this research is to explore the benefits of using a software analytics platform for practitioners managing quality.
\textit{Method:} In 
a technology transfer project, a software analytics platform was incrementally developed between academic and industrial partners to address their software quality problems. 
This paper focuses on exploring the value provided by this software analytics platform in two pilot projects.
\textit{Results:} Practitioners emphasized major benefits including the improvement of product quality and process performance and an increased awareness of product readiness. They especially perceived the semi-automated functionality of generating quality requirements by the software analytics platform as the benefit with the highest impact and most novel value for them.
\textit{Conclusions:} Practitioners can benefit from modern software analytics platforms, especially if they have time to adopt such a platform carefully and integrate it into their quality assurance activities. 

\end{abstract}

\begin{IEEEkeywords}
software quality, software engineering, software analytics, technology transfer, summative evaluation
\end{IEEEkeywords}

\section{Introduction}
Software analytics is about utilizing data-driven approaches to obtain insightful and actionable information to help software practitioners with their data-related tasks \cite{gall2014software}. Software analytics platforms provide features for analyzing and visualizing software development data to support data-driven decision-making \cite{Zhang2013, martinez-fernandez2019sat}. Despite the growth in the use of software analytics platforms in industry, little empirical evidence is available about the extra value they provide \cite{Berntsson19}.

Our research goal is to explore the benefits of using a software analytics platform for practitioners managing quality. We applied technology transfer \cite{Gorschek2006, pfleeger1999understanding, Diebold2015}, in close collaboration and cooperation between practitioners and researchers in the Q-Rapids research and innovation project. We use the technology transfer definition of Bandyszak et al.: ``the process of sharing or developing a technology object between two or more actors via one or more media so that the technology recipient sustainably adopts the object in the recipient’s context in order to evidently achieve a specific purpose.'' \cite{Bandyszak2016}.
We report the industrial scenario and describe our experiences on the industry-academia journey in two pilot projects that used a software analytics platform to solve their quality management problems. Our previous work focused on incrementally developing a software analytics platform in a previous formative stage \cite{Lopez2018}. This study reports our experiences with the software analytics platform Q-Rapids (hereafter referred to as Q-Rapids) during its summative evaluation. It is available at https://github.com/q-rapids.

In Section II, we describe the context of our study, including the software analytics platform and our two companies. Section III contains the research methodology applied, while Section IV presents the challenges and benefits experienced while using Q-Rapids. Section V contains threats to validity and Section VI presents related work. The paper concludes with a summary and next steps in  Section VII.



\section{Research setting}
This section describes the software analytics platform used in the two companies that took part in this study. To fully understand the context of the dynamic validation (i.e., pilot projects) \cite{Gorschek2006} in the companies, we report their application domain, problem statements, needs in terms of managing quality, and how the software analytics platform was introduced.

\subsection{The software analytics platform}
The software analytics platform Q-Rapids was developed as part of the Q-Rapids project\footnote{https://www.q-rapids.eu/} and provides tool-supported quality management in the context of agile software development \cite{franch17, Martinez18, Lopez2018}. It includes several software analytics capabilities to support data-driven decision-making, such as collecting and integrating data from different data sources, real-time modeling of this data, prediction and simulation analysis, and suggestions of quality requirements that can be semi-automatically transferred to the company's product backlog system (see Fig. \ref{open-project}). Q-Rapids provides a dashboard to support the usage of all these data-driven decision-making capabilities.

\subsection{Company A}
\textbf{Application domain.} Company A specializes in the development of reliable, secure communications and connectivity solutions and offers proven information security solutions for mobile devices and portable computers. The company provides innovative products and services, customized solutions based on its product platforms, and R\&D services. Company A 
also provides healthcare technology products and services for the medical domain.

\textbf{Problem statement and quality management goals.} Company A has been looking for enhancing their software-development-focused data analysis and finding fast feedback methods regarding functional and non-functional requirements, i.e., quality requirements. In addition, the purpose has been to enhance a holistic view on system, product, or operations level by using data mining and analysis. 
These objectives are addressed by the utilization of Q-Rapids dashboards and predictive analysis. These features enable decision-making regarding readiness or performance related to functional and non-functional requirements, which is a very attractive application from a company point-of-view.\\
Company A specified the following expected impact indicators from using Q-Rapids:
\begin{itemize}
    \item \textit{Feature throughput}. Increased percentage of features that meet time-to-market targets with the desired levels of quality.
    \item \textit{Release frequency}. Increased number of releases per time unit.
    \item \textit{Realized requirements}. Increased proportion of quality requirements that are used in actual features and releases. These factors can be compared to the overall number of the requirements.
    \item \textit{Product quality}. Improved product quality, referring to the maintainability, reliability, and functional suitability of the software product. These address the non-functional requirements of the product being developed.
    \item \textit{Process performance}. Improved process performance regarding the software development in terms of efficiency and quality of the software lifecycle processes used.
\end{itemize}

\textbf{Use of the software analytics platform in a pilot project.} The alignment of the company's software development process and the company tool chains together with the configuration of Q-Rapids has been an important factor in building a successful implementation of Q-Rapids. The first implementation of Q-Rapids took place in 2018 and it has been up and running for nearly one year \cite{martinez-fernandez2019sat}.
The second pilot consists of a product family with five products and related components in the context of product information systems development. Thus, the establishment involved a larger set-up with a total of five deployments of Q-Rapids. The team of this project consists of a product owner, a project manager, and some 15 developers.
In addition, the platform has now been disseminated widely within the company, arousing great interest.
\subsection{Company B}
\textbf{Application domain.} Company B is a tool vendor developing a 25-year-old product line of a model-driven tool suite dedicated to expressing and managing requirements, modeling software architectures, building accurate  models, and automating application code production for several languages.

\textbf{Problem statement and quality management goals.} Considering that these products are used by its customers to develop critical systems in the military and transportation domains, company B is committed to providing a quality guarantee for the software it releases. To achieve this goal, company B expressed the need to evolve the quality management processes applied to its software development cycle by getting new insights to support the decision-making process in the context of rapid software development, to automate the management of the quality requirements process across the organization, and to centralize the heterogeneous data sources related to product quality in a software analytics platform.\\
Improving the quality of company B's products involves monitoring and improving several key indicators:
\begin{itemize}
\item \textit{Product quality}. Refers to the maintainability, reliability, and functional suitability of the software product.
\item \textit{Product readiness}. Refers to a product that is “ready to be released” (i.e., a product implementing the features planned in the release and without blocking issues).
\item \textit{Quality feedback loop}. Refers to the proportion of quality requirements that are used in actual features and releases.
\item \textit{Productivity rate}. Refers to the time used for the development and testing of new features / time used for maintenance or defect removal.
\end{itemize}

\textbf{Use of the software analytics platform in a pilot project.} Company B deployed Q-Rapids in the development team of its flagship product and used the platform to monitor two release cycles of this product over one year. By collecting data related to code analysis, issue management, project management, integration and testing, and software configuration management, company B was able to aggregate indicators covering the whole of its software development process. Next, a selection of team members occupying several roles, such as product owner, project manager, quality engineer, and developer, were trained in the use of Q-Rapids. In addition, they worked on the integration of Q-Rapids into their development process.
\begin{figure}[htbp]
\centerline{\includegraphics[width=0.5\textwidth]{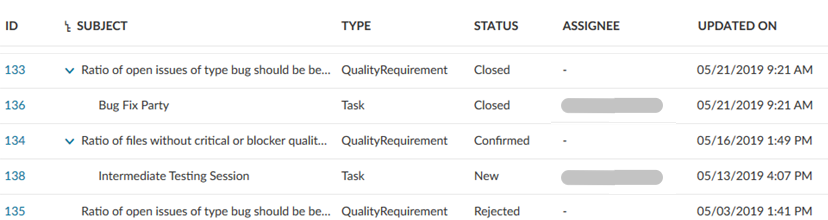}}
\caption{Screenshot of the product backlog of the company containing quality requirements generated by Q-Rapids.}
\label{open-project}
\end{figure}
\section{Research methodology}
The following subsections describe the (a) research goal and question, (b) research design, (c) population and sampling, (d) instruments, (e) execution, and (f) data analysis of our study.

\subsection{Research goal and question}
This study focuses on exploring the application of the software analytics platform Q-Rapids under realistic circumstances and related benefits for managing quality. Therefore, we defined our goal following the GQM approach \cite{gqm94} as: Analyze the impact and related benefits of Q-Rapids with respect to realistic usage from the perspective of project managers, technical leads, and developers in the context of managing quality in agile software development. This led to the following research question: \textit{Which value does a software analytics platform provide in 
the two pilot projects under study?}
%

\subsection{Research design}
To answer the research question, we followed a technology transfer approach in two companies \cite{Gorschek2006}. 
This approach distinguishes between two phases: formative and summative. First, the formative stage focuses on supporting the evolution of concepts and ideas mainly from research work. Thus, we evaluated the first prototype of Q-Rapids 
as well as the intermediate version containing more functionalities   
with industry partners at the beginning and end of 2018 in controlled environments \cite{d5.1}. After the formative stage, we started with the integration of Q-Rapids in real settings in order to validate the software analytics platform with practitioners actually using it (summative stage) \cite{Easterbrook2008}. This summative stage is the focus of this study. 
Thus, we planned a research study mainly consisting of two parts: \begin{enumerate}
    \item Tracking the use of the software analytics platform over a period of time and reacting, if necessary, and 
    \item Collecting feedback from practitioners on-site in real pilot projects.
\end{enumerate}
%
%

\subsection{Population and Sampling}
The target population included all the practitioners from both companies who had been working with Q-Rapids and could therefore report their experiences to us. The main roles we expected were product owner, manager, technical lead, and developer.
We contacted the companies and asked for suitable persons to draw a convenient sampling \cite{daniel2011sampling}.

In total, 13 people participated in the evaluation workshop (see Table \ref{demographics}) including five managers, three developers, one technical lead, one team lead, one engineer, and one person from company A did not specify the role.
\begin{table*}[htbp]
\caption{Demographic information of participants.}
\begin{center}
\begin{tabular}{|c|c|c|c|c|c|c|c|c|c|}
\hline
& \textbf{Participants} & \textbf{Roles} & \textbf{Work experience} & \textbf{Work experience} & \multicolumn{5}{|c|}{\textbf{Frequency of using Q-Rapids}} \\
& & & \textbf{in company} & \textbf{in role} & \multicolumn{5}{|c|}{} \\ 
& & & (min - max) & (min - max) & 2-5 times & monthly & bi-weekly & weekly & daily\\ 
\hline
\textbf{Company} & 8 & 3x Project Manager & 5.5 years & 2.5 years & 3 & 1 & 1 & 2 & 1 \\
\textbf{A}& & 2x Technical Lead & (6 months - 20 years) & (1 year - 7 years) &  &  &  &  &  \\
& & 2x Developer &  &  &  &  &  &  &  \\
\hline
\textbf{Company} & 5 & 2x Project Manager & 11 years & 7 years & 1 & 0 & 2 & 2 & 0 \\
\textbf{B} & & 1x Developer & (2 - 14 years) & (1 year - 10 years) &  &  &  &  & \\
 & & 2x Other & & & & & & & \\
\hline
\end{tabular}
\label{demographics}
\end{center}
\end{table*}

\subsection{Data collection instruments}
Our research design mainly consisted of two different parts for collecting relevant data.
First, we developed the online questionnaire to track the usage of the software analytics platform by implementing a survey in Limesurvey\footnote{https://www.limesurvey.org/}. The practitioners were asked to report on their real usage on a monthly basis at least. This survey asked for a \textit{usage story} to explain how the software analytics platform was being used.
Answers were supposed to look like this: \textit{As a} product owner, \textit{I have used} the quality assessment of Q-Rapids, \textit{to} monitor the issue resolution \textit{during} sprint grooming.
Furthermore, the survey asks for strengths and suggestions for improvements related to the defined usage.

The other part of the method triangulation for collecting data was the evaluation workshop. The procedure of this evaluation workshop consisted of four different sessions:
\begin{enumerate}
    \item Introduction to the evaluation objectives and procedures including the signing of the informed consent and filling out demographic information.
    \item Presentation of the company-specific context and information about previously defined impact indicators of the software analytics platform based on the quality management goals of Section II.
    \item Collection of feedback (1) using a structured questionnaire for collecting the individual perceptions and (2) during a dynamic group discussion part, where the previously made prioritization of the impact indicators had to be validated and more detailed discussions about the added value of using the software analytics platform were moderated. Therefore, the participants were asked to form several groups, which were asked to fill out at least one template with respect to one of the most relevant impacts. Afterwards, the results of each group were presented to the other participants.
    \item Closing of the evaluation workshop with a brief summary and presentation of the next steps.
\end{enumerate}
The materials used during the evaluation are available online\footnote{https://figshare.com/collections/Q-Rapids/4584038} (except for the company-specific presentations about the available data measuring impact indicators and their current values): (1) a questionnaire asking about the perceived impact of using the software analytics platform, i.e., the participants had to individually rate how the software analytics platform impacted their work with respect to the predefined and expected beneficial indicators and which benefits they observed; (2) a template for the moderated group dynamic session that focused on details about a particular impact and the maturity of the corresponding implementation.

\subsection{Execution}
We trained members of both companies in the usage of Q-Rapids in November 2018 and demonstrated each feature using their own collected data. Afterwards, the system was still running in the companies and the people could start using the platform in their daily work.

After a short period of familiarization with Q-Rapids in practical settings, we activated the online questionnaire in mid-January 2019 and received the first feedback about the usage of Q-Rapids in real environments at the end of January.
The evaluation workshops in both companies were executed in parallel at the beginning of June 2019 following the predefined evaluation protocol. Each of the first two authors moderated one workshop and was supported by other researchers who acted as observers. The workshops had a duration of up to 3 hours including a break.

\subsection{Data analysis}
The designed instruments allowed us to collect quantitative as well as mainly qualitative information from the practitioners. Therefore, we started analyzing the qualitative data using a thematic analysis \cite{Braun06:TA}. We used the data collected by the online questionnaire to report on the challenges the practitioners had encountered while using the software analytics platform under real circumstances. In addition, we performed a qualitative analysis based on the data collected during the evaluation workshop by using the completed impact templates as the main basis to derive and categorize the results. We extended this written data with all the information obtained through the group discussions documented in the observation protocols. To obtain a ranking of benefits according to their impact, we used the quantitative data from our questionnaire distributed during the workshop session. We computed descriptive statistics including the sample size, minimum, maximum, median, mode, and average to distinguish between the ratings of the participants. This analysis will be conducted again after collecting more data to perform a more powerful interpretation of our results (thus, we will report quantitative results in the next publication).

\section{Preliminary Results in two Companies}
Based on our research question, we report the main results of our data analysis from the information collected with the different instruments:
\subsection{Challenges}
During the deployment as well as during the usage of the software analytics platform, both companies faced similar challenges. We reported a few of them on \cite{martinez-fernandez2019sat}, e.g., \textit{need for tailoring to the company}, \textit{need for integration with other tools}, \textit{simplification of platform installation}, \textit{need for an efficient configuration process}. According to the feedback of the real usage collected through the online questionnaire, especially the challenges with the set-up and configuration of Q-Rapids made it difficult for the participants to distinguish between configuration changes, bugs of the system, and changes of people's behavior when interpreting historical trends of the software analytics results. Furthermore, they encountered bugs during usage that had been fixed during the last months to correctly calculate the analyses and to convince colleagues to start using the platform.

\subsection{Benefits}
\subsubsection{Company A}
All participants of company A considered three main benefits as positively impacted by Q-Rapids (ordered by perceived impact):

1. Generation of quality requirements by Q-Rapids is a novel and important functionality for the company. Getting alerts to ensure that the quality requirements are maintained throughout the project lifecycle and the possibility to predict and simulate the quality were perceived as great features valued by the participants. Data-driven decision-making reduces manual "guess work", and at present none of the currently used tools provides this kind of functionality. In order to monitor the status of the realized requirements, the company tracks which requirements were created using Q-Rapids and are set to \textit{close} in their backlog system.

2. Monitoring the process performance based on the measured data helps to reflect on and check the process itself. Therefore, the company has defined the development speed, testing performance, and issue velocity as key aspects of their process quality monitoring and uses several available data sources, such as their issue-tracking system Jira\footnote{https://www.atlassian.com/software/jira}, to collect the necessary data. One of the evaluation participants explained that they tried to also follow up on the number of story points in relation to development speed. So, the current benefit of Q-Rapids could be improved even more if more effort were applied to enhance the precision of the currently used quality model and complement it. In addition, the participants highlighted the traceability of changes over time, i.e., they reasoned about identified change points in order to better understand their processes (e.g., whether the value was low during the time of somebody's vacation). Now the process-related quality can be measured efficiently.

In addition, communication within the team has increased and different roles are connected within the same environment, where several tools and data sources are combined; e.g., product owners - project managers - developers - testers.

3. The consideration of product quality during development is supported by Q-Rapids in such a way that it provides a data-driven baseline for decision-making. Therefore, Q-Rapids collects the quality-related data from multiple sources and visualizes all information in one place. The company is currently using SonarQube\footnote{https://www.sonarqube.org} to improve software quality, but ideally, the same (and more) could be done with Q-Rapids. Furthermore, the usage of the prediction and simulation features improves the decision-making process.

In addition, the participants explained that the \textit{Release frequency} would be more suitable and relevant if the focus were switched to release accuracy instead of frequency, i.e., "\textit{Release frequency} is important but release accuracy is more important. The purpose would be to be able to build new releases based on a healthy baseline" within the intended time boxing instead of a standard release every day.

\subsubsection{Company B}
The use of the Q-Rapids platform has brought four main benefits to company B, ordered by their impact from the perspective of the team:

1. Consideration of semi-automated quality requirements is integrated into the software process. By using Q-Rapids, the accepted quality requirements (those suggested by Q-Rapids and accepted by the Q-Rapids user) are automatically added to the product backlog of the company, where they are assigned to team members (see Fig. \ref{open-project}). In addition, two key metrics are created and are monitored to measure the success of this activity: the fulfillment of these quality requirements (i.e., when they are solved and closed by the assignee), and their relevance.

2. Improvement and transparency regarding product quality. Q-Rapids has been offering a centralized repository with all relevant metrics with respect to the software quality of one of the products of company B (e.g., code quality metrics and bug reports). The participants mentioned that it has been helpful to be aware of how product quality can be improved and where to invest more effort based on the simulation and prediction functionalities. Besides improving the quality of their product, they remarked that this brought transparency regarding how software quality is viewed, which is sometimes hidden due to the time pressure of delivering new features and adding value to the product. 

3. Having an indicator integrating many aspects of the process performance monitored and visualized within Q-Rapids. Due to the data coming from heterogeneous data sources, such as test-related data from continuous integration systems (e.g., Jenkins\footnote{https://jenkins.io/}) and the proprietary testing tool for integration testing, this indicator provides a new overview of the aggregated status of the testing process during the development process.

\begin{figure}[htbp]
\centerline{\includegraphics[scale=0.31]{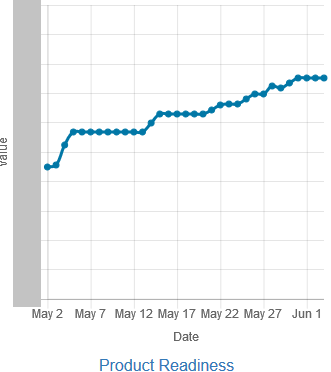}}
\caption{Screenshot of Q-Rapids displaying the indicator \textit{Product Readiness}.}
\label{dashboard}
\end{figure}

4. Increased awareness of product readiness. Company B has created another indicator called product readiness in Q-Rapids (see Fig. \ref{dashboard}) calculated on the basis of data from different lifecycle phases (development, tests, builds). Therefore, the participants stated that when using Q-Rapids, they were aware of when the product could be released, respectively whether a release was likely to be postponed. An additional related benefit is that the aforementioned factors from different phases are the responsibility of different teams, so the creation of alerts for each sub-factor/team could improve communication among teams.


\section{Threats to Validity}
\textit{Construct Validity}. We combined and used different data collection instruments. Moreover, we included open questions and comment fields in our questionnaires to encourage the participants to provide more explanations for their opinions.

\textit{Conclusion Validity}. All the collected data was analyzed to cross-check the results in order to draw reliable conclusions. In addition, we combined qualitative and quantitative analysis. Afterwards, we validated these results with our previous ones from the formative evaluations, and both companies approved their corresponding results. 

\textit{Internal Validity}. We conducted both evaluation workshops in parallel due to the availability of the participants. Thus, the workshops were moderated by two different persons. In order to conduct both workshops in a  similar manner, the moderation followed a predefined evaluation protocol. In addition, the intensity of using Q-Rapids differed among the participants, which might have influenced their assessment of the impact. As we aimed to collect honest and open-minded feedback from the practitioners, we assured them of the anonymity of the data they provided by means of an informed consent.  

\textit{External Validity}. The results of this study are only based on the context of two companies and we can therefore not generalize our findings to other companies or even to other settings.

\section{Related Work}
Recent studies report experiences and impact factors regarding the use of software analytics platforms.

Izquierdo et al. developed a software development analytics platform for Xen \cite{izquierdo2019}. The main benefit was an increased understanding of the time to merge (the time that a change is under code review, from the moment it is proposed to the moment it is eventually accepted into the code base). For Xen's stakeholders, higher time to merge negatively affects the capability to deliver new features to customers.


Huijgens et al. studied factors helping the application of software analytics in ING (e.g., dashboard containing only a limited number of metrics and infrastructure for building dashboards) and hindering it (e.g., dashboard is not used by the squad or is not user-friendly) \cite{Huijgens2018}. In the same company, they used their software analytics platform to study timing and quality characteristics of rapid releases \cite{kula2019}.

Svensson et al. report that in a survey of 84 practitioners, the practitioners were positive about the future use of data-driven decision-making systems for higher-level and more general decision-making, fairly positive about its use for requirements elicitation and prioritization decisions, and less positive about its future use at the team level \cite{Berntsson19}.

Our work focuses on the challenges and value of using a software analytics plaform, including advanced techniques such as quality requirements generation, assessment and improvement of software quality, and following the performance of the software development process.

\section{Conclusion and next steps}
In this study, we investigated experiences gathered from the use of the software analytics platform Q-Rapids in two companies. At the time of this study, the development of Q-Rapids had arrived at its summative stage, meaning that the industry partners had started to use Q-Rapids in real settings in selected projects. 
The challenges faced during the set-up and usage of Q-Rapids were similar in both companies. For example, the configuration is effort-intense and might influence the reliability of the software analytics results, which makes it hard for practitioners to distinguish between buggy results and a real quality shift of their project.
Yet based on the conducted work, Q-Rapids now provides a comprehensive overview of the quality status of the pilot projects. 
Thus the value provided by the software analytics platform to the companies includes the novel functionality to semi-automatically generate quality requirements to support quality management, the monitoring of product quality and  process performance, and increased awareness of product readiness.

Based on these results, company A plans to further improve the stability of Q-Rapids and "package it" for easy deployment. After that, they can start spreading it for use in other projects and by other persons. They also expect to polish their processes with more data-driven decision-making based on the information provided by Q-Rapids.

Company B is planning to further integrate the use of Q-Rapids into the product by all the stakeholders (instead of merely a subset of members). They predict an increase in the number of quality requirements semi-automatically managed by Q-Rapids, to solve bugs due to low code quality and poor tests before production. Furthermore, they plan to fully integrate the usage of Q-Rapids into the development process (e.g., during release planning) and to scale Q-Rapids as part of the development chain.


\section*{Acknowledgment}
The authors thank all the members of Bittium and Softeam who participated in the evaluation. They also thank Prabhat Ram and Woubshet Behutiye for supporting the workshops onsite and Sonnhild Namingha for proofreading. This work was supported by the European Union's Horizon 2020 Research and Innovation Programme under grant agreement 732253 (Q-Rapids: Quality-Aware Rapid Software Development).

\bibliographystyle{./IEEEtran}
\bibliography{./IEEEabrv,./Paper-Bibliography}

\end{document}